\documentclass[twocolumn,aps,pra,superscriptaddress,floatfix,nofootinbib]{revtex4-2}
\usepackage[latin9]{inputenc}
\usepackage{color}
\usepackage{amstext}
\usepackage{graphicx}
\usepackage[colorlinks, allcolors={blue}]{hyperref}
\usepackage{physics}
\usepackage{ulem, cancel}
\usepackage{orcidlink}

\usepackage{xcolor,graphicx}
\usepackage{amssymb,amsmath,mathrsfs}
\usepackage{mathtools}
\usepackage{verbatim}
\usepackage{bbm}
\usepackage{multirow}

\newcommand{\stkout}[1]{\ifmmode\text{\sout{\ensuremath{#1}}}\else\sout{#1}\fi}

\makeatletter
\def\p@subsection{}
\def\p@subsubsection{}
\makeatother

\begin{document}

\title{Unveiling quantum complementarity tradeoffs in relativistic scenarios}

\author{Marcos L. W. Basso \orcidlink{0000-0002-0416-3582}}
\email{marcoslwbasso@hotmail.com}
\affiliation{Center for Natural and Human Sciences, Federal University of ABC, Avenida dos Estados 5001, 09210-580, Santo Andr\'{e}, S\~{a}o Paulo, Brazil}

\author{Ismael L. Paiva \orcidlink{0000-0002-0416-3582}}
\email{ismaellpaiva@gmail.com}
\affiliation{H. H. Wills Physics Laboratory, University of Bristol, Tyndall Avenue, Bristol BS8 1TL, United Kingdom}

\author{Pedro R. Dieguez \orcidlink{0000-0002-8286-2645}}
\email{dieguez.pr@gmail.com}
\affiliation{International Centre for Theory of Quantum Technologies, University of Gda\'nsk, Jana Bazynskiego 8, 80-309 Gda\'nsk, Poland}

\begin{abstract}
Complementarity plays a pivotal role in understanding a diverse range of quantum phenomena. Here, we show how the tradeoff between quantities of a complete complementarity relation is modified in an arbitrary spacetime for a particle with an internal spin. This effect stems from local Wigner rotations in the spacetime, which couple the spin to the system's external degrees of freedom. To conduct our study, we utilize two generalized delayed-choice interferometers. Despite differences in complementarity tradeoffs inside the interferometers, the interferometric visibility of both setups coincides in any relativistic regime. Our results extend the finding that general relativity induces a universal decoherence effect on quantum superpositions, as local Wigner rotations, being purely kinematical, preclude any spin dynamics. To illustrate, we analyze the Newtonian limit of our results.
\end{abstract}

\maketitle

\section{Introduction}
\label{sec:introduction}

Bohr's complementarity principle suggests that physical systems may have properties that only emerge once the entire physical context for their probing is settled~\cite{bohr1928quantum, bohr1935can}. This idea sparks debates to this date~\cite{jackes2007experimental, manning2015wheeler, kastner2017beyond, vedovato2020extending, catani2021interference, catani2022aspects}, with various theoretical approaches attempting to quantify it~\cite{wootters1979complementarity, greenberger1988simultaneous, englert1996fringe, durr2001quantitative, saunders2005complementarity, englert2008wave, baumgratz2014quantifying, bera2015duality, angelo2015wave, coles2016entropic, bagan2016relations, bagan2018duality, mishra2019decoherence, basso2020quantitative, basso2020complete, basso2021uncertainty, qureshi2021predictability, basso2022entanglement2}. Besides foundational debates, it is also known that complementarity is needed for the achievability of some tasks, e.g., it is a resource for unambiguous exclusion and encryption~\cite{hsieh2023quantum}.

This work follows recent efforts to investigate the interplay between gravity and quantum mechanics and to characterize general relativistic effects in quantum phenomena~\cite{terashima2004einstein, fuentes2005alice, zych2011quantum, zych2012general, martin2014entanglement, pikovski2015universal, margalit2015self, zych2016general, bose2017spin, marletto2017gravitationally, christodoulou2019possibility, howl2019exploring, roura2020gravitational, basso2021interferometric, basso2021effect, nemirovsky2022spin, belenchia2022quantum, danielson2023killing}. We consider, in particular, the case of massive particles with an internal spin. Our objective is twofold. First, we aim to understand how complementarity is affected by a general spacetime structure, a fundamental result that can serve as a basis for modifications of protocols that make use of this concept. Second, we aim to analyze whether an existing result by Zych \textit{et al.}~\cite{zych2011quantum} for spinless particles continues to hold (in leading order) for spin particles. The latter work showed how the visibility of an interferometer is affected by general relativistic effects if a spinless particle with an internal clock transverses it. However, for spin particles, Wigner rotations~\cite{silberstein1914theory, thomas1926motion, wigner1939unitary} have to be taken into account, as we will discuss in more detail.

In our investigation, we make use of complete complementarity relations (CCRs)~\cite{jakob2010quantitative, basso2020complete}. In particular, to study complementarity \textit{inside} an interferometer, where path coherence (and not visibility~\cite{bera2015duality, bagan2016relations, dieguez2022experimental, chrysosthemos2023quantum, wagner2022coherence}) can be taken as a good measure of nonclassicality, we consider the CCR~\cite{basso2020complete}
\begin{equation}
    \label{eq:ccr}
    C_{\text{re}}(\rho_S) + P_{\text{vn}}(\rho_S) + S(\rho_S) = \log_2 d_S,
\end{equation}
where $C_{\text{re}}(\rho_S)$ is the path relative entropy of coherence, $P_{\text{vn}}(\rho_S)$ is the von Neumann predictability, and $S(\rho_S)$ is the von Neumann entropy associated with a system $S$ in the state $\rho_S$.

The coherence $C_{\text{re}}(\rho_S) \coloneqq S(\rho_S || \Phi_{Z}(\rho_S))$, where $\Phi_Z(\rho_S) \coloneqq \sum_\eta Z_\eta \rho_S Z_\eta$ denotes the output of a non-selective measurement of $Z=\sum_\eta z_\eta Z_\eta$ with projectors $Z_\eta = \ket{\eta}\bra{\eta}$ associated with paths of the interferometer, captures a wave-like behavior of the system. The predictability $P_{\text{vn}}(\rho_S)\coloneqq S(\Phi_Z(\rho_S)||\Phi_{ZX}(\rho_S)) = S^{\text{max}} - S(\Phi_Z(\rho_S))$, where $X$ is a discrete-spectrum observable whose eigenbasis is mutually unbiased to the eigenbasis of $Z$, measures how much the probability distribution associated with $\Phi_Z(\rho_S)$ differs from the uniform probability distribution $I_\mathcal{S}/d_\mathcal{S}$. It can be interpreted as \textit{a priori} path information~\cite{greenberger1988simultaneous}, capturing a particle-like behavior of the system. Finally, $S(\rho_S)$ can be thought of as the relative entropy of entanglement between system $S$ and its purification. These correlations store which-path information about the external degrees of freedom of the system, also leading to a particle-like behavior. It should be noted that, despite the long discussion about entanglement in relativistic scenarios~\cite{czachor1997einstein, gingrich2002quantum, peres2002quantum, terashima2003relativistic, terashima2004einstein, peres2004quantum, lamata2006relativity}, the CCR in Eq.~\eqref{eq:ccr} has been shown to be Lorentz invariant~\cite{basso2021complete, basso2021complete2, basso2021complete3}.

\begin{figure*}
    \centering
    \includegraphics[width=\textwidth]{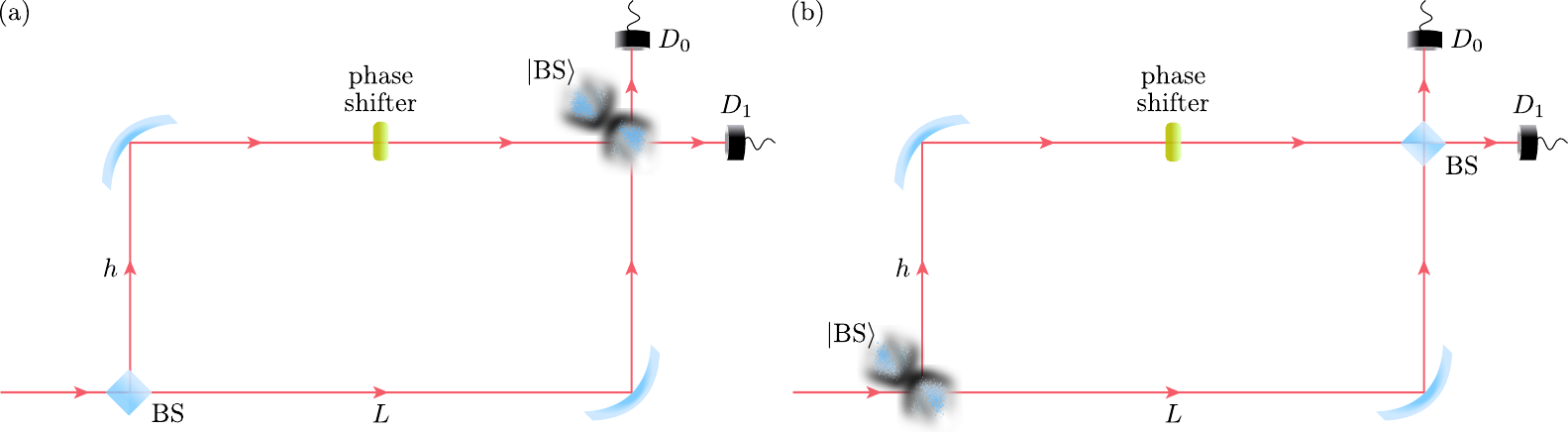}
    \caption{Schematic representation of the delayed-choice experiments. (a) In the quantum delayed-choice experiment (QDCE), proposed in Ref.~\cite{ionicioiu2011proposal}, the system enters an interferometer with an unbiased beam splitter (BS) and a phase shifter that implements a relative phase between the two paths. A second BS is considered in a coherent spatial superposition of being {\it in} and {\it out} of the interferometer. (b) In the quantum-controlled reality experiment (QCRE), introduced in Ref.~\cite{dieguez2022experimental}, the experimental arrangement is based on swapping the BSs of the QDCE. The first BS is now taken to be in a quantum coherent superposition of being in and out of the interferometer. It is noteworthy that, besides this difference, both experimental arrangements lead to the same interferometric visibility.}
    \label{fig:interferometers}
\end{figure*}

The interferometers considered in our analysis are two variations of Wheeler's delayed-choice experiment~\cite{wheeler1978past, wheeler1983law}, known as quantum delayed-choice experiment (QDCE)~\cite{ionicioiu2011proposal} and quantum-controlled reality experiment (QCRE)~\cite{dieguez2022experimental}. QDCE consists of a Mach-Zehnder in which the BS at the end of the interferometer is prepared in a quantum superposition of being present and absent in the interferometer path, as represented in Fig.~\ref{fig:interferometers}(a). Observe that the BS at the beginning of the interferometer is standard. This interferometer attracted attention because its output presents statistics that seem to be a wave-and-particle hybrid~\cite{adesso2012wave, auccaise2012experimental, roy2012nmr, peruzzo2012quantum, kaiser2012entanglement}. However, its precise consequence to the understanding of the complementarity principle is still up for debate (see, e.g., Refs.~\cite{coles2014equivalence, angelo2015wave, chaves2018causal}).

QCRE, on the other hand, modifies the QDCE to ensure that systems traveling inside the interferometer have a hybrid wave-and-particle behavior from the perspective of an entropic measure dubbed realism~\cite{bilobran2015measure,dieguez2018information}. To achieve this, the positions of the standard BS and the one prepared in a superposition were exchanged, as illustrated in Fig.~\ref{fig:interferometers}(b). While this interferometer has the same visibility as the QDCE, it establishes a monotonic relation between the output visibility with the wave and particle-like behavior of the system~\cite{dieguez2022experimental}.

It is worth highlighting that, upon introducing a new quantum system that can, in principle, be measured even after the interferometer is encircled (like the superposed BS), problems related to the measurement problem can be evoked. However, solutions (and even formulations) of the measurement problem often rely on a chosen interpretation of the theory. Since these choices are indistinguishable at the operational level, our analysis focuses on how gravity influences the tradeoff between operationally significant quantities.

We study the CCR for the QDCE and QCRE in a nonrelativistic scenario in Sec. \ref{sec:preliminaries}. This serves as a basis for comparison with our first results in Sec.~\ref{sec:grtreat}, which deal with these interferometers embedded in a general spacetime. Next, we make our analysis more concrete with an investigation of the Newtonian limit of our results in Sec.~\ref{sec:new-lim}. To conclude, we present our final discussion and outlook in Sec.~\ref{sec:discussion}. In the Appendices, we briefly present the relevant technical background to this work, mathematical derivations, and other conceptual discussions.

\section{Preliminaries}
\label{sec:preliminaries}

Before studying the relativistic treatment of QDCE and QCRE, we first present a nonrelativistic analysis of these experiments, since they had not yet been investigated with the CCR in Eq.~\eqref{eq:ccr} in the literature.

We start with the so-called quantum version of Wheeler's delayed-choice experiment (QDCE) proposed in Ref.~\cite{ionicioiu2011proposal} and depicted in Fig.~\ref{fig:interferometers}(a). Denoting $\{0,1\}$ as the paths to be traveled in the interferometer and $\ket{\psi_{\text{i}}}=\ket{0}\otimes\ket{\text{BS}}$ as the initial state, with $\ket{\text{BS}} \equiv \cos{\alpha} \ket{\text{in}} + \sin{\alpha} \ket{\text{out}}$ representing a beam splitter (BS) in a coherent superposition of being present (in) and absent (out) of the interferometer with $0\leq \alpha \leq \pi/2$. The joint system can be described right after the phase shifter introduced in path $1$ as
\begin{equation}
    \label{eq:psi_WDCE}
    \ket{\psi_{\text{QDCE}}} = \frac{1}{\sqrt{2}} \left(\ket{0} + e^{i\phi}\ket{1}\right) \ket{\text{BS}}.
\end{equation}
At this stage, this is a separable state that does not yet contain any correlation between the particle and the BS in superposition. In fact, the state is unchanged until right before the system passes through the BS.

With this, we can readily construct the CCR in Eq.~\eqref{eq:ccr} for a system in this region of the interferometer. Indeed, it is straightforward that predictability and entanglement vanish for the state in Eq.~\eqref{eq:psi_WDCE}, while coherence between the paths is maximum. Hence, there is no tradeoff between the different quantities of the CCR, a characteristic similar to that observed in Ref.~\cite{dieguez2022experimental} from the realism perspective.

We can also compute the visibility associated with this configuration. Indeed, observe that the output state right after the BS in superposition is
\begin{equation}
    \begin{aligned}
        \ket{\psi_{\text{f}}}= &\frac{1}{\sqrt{2}} \cos{\alpha} \left(\ket{x+} + e^{i\phi} \ket{x-}\right) \ket{\text{in}} \\
            &+ \frac{1}{\sqrt{2}} \sin{\alpha} \left(\ket{0} + e^{i\phi} \ket{1}\right) \ket{\text{out}},
    \end{aligned}
\end{equation}
where $|x\pm\rangle = (|0\rangle \pm |1\rangle)/\sqrt{2}$ is written in the standard Pauli notation. The probability distribution for the detector $D_0$ at the output of the interferometer reads
\begin{equation}
    \mathfrak{p}_0=\frac{1}{2}\left(1 + \mathcal{V} \cos{\phi}\right),
\end{equation}
where 
\begin{equation}
    \mathcal{V} \coloneqq \frac{\mathfrak{p}_0^{\text{max}}-\mathfrak{p}_0^{\text{min}}}{\mathfrak{p}_0^{\text{max}}+\mathfrak{p}_0^{\text{min}}}= \cos^2\alpha
\end{equation}
is the final interferometric visibility.

Now, let us consider the quantum-controlled reality experiment (QCRE) proposed in Ref.~\cite{dieguez2022experimental}, which is schematically represented in Fig.~\ref{fig:interferometers}(b). Again, we assume the initial state of the joint system to be the same $\ket{\psi_{\text{i}}}$ as before. However, $|\text{BS}\rangle$ refers to the state of the first BS this time around. Then, after the BS in superposition and the phase shifter on path $1$, the joint system can be described as
\begin{equation}
    \label{phi-QCRE}
    \ket{\phi_\text{\tiny QCRE}} = \frac{1}{\sqrt{2}} \cos{\alpha} \left(\ket{0} + e^{i\phi} \ket{1}\right) \ket{\text{in}} + \sin{\alpha} \ket{0} \ket{\text{out}}.
\end{equation}
Again, this state is kept unmodified until right before the second (standard) BS.

\begin{figure}
    \centering
    \includegraphics[width=1\columnwidth]{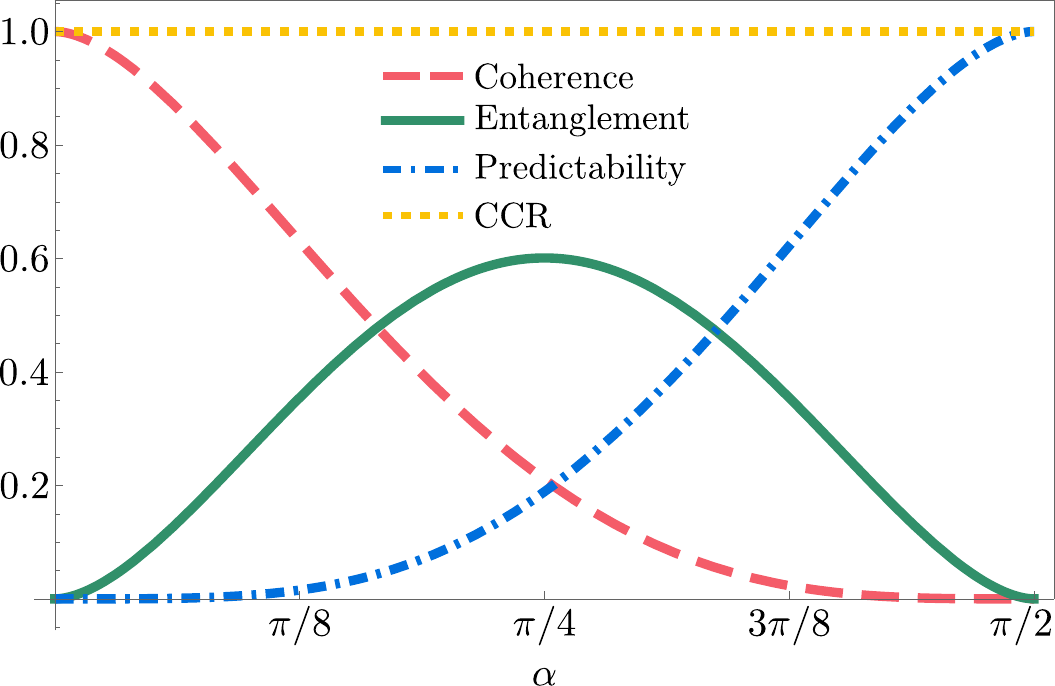}
    \caption{CCR for the nonrelativistic QCRE as a function of $\alpha$. The plot shows how coherence and predictability follow a monotonic relation with the angle $\alpha$, while the entanglement is maximum when the quantum controller has maximal initial coherence.}
    \label{fig:nr-qcre}
\end{figure}

We can then compute the CCR associated with $\ket{\phi_\text{\tiny QCRE}}$. First, we trace out the BS to obtain the reduced density matrix of the system traveling the interferometer $\rho_S = \text{Tr}_{\text{BS}}\ket{\phi_\text{\tiny QCRE}} \bra{\phi_\text{\tiny QCRE}}$, which gives
\begin{equation}
    \rho_S = \frac{1}{2}\left(
    \begin{array}{cc}
        2-\cos^2\alpha &  e^{-i\phi} \cos^2\alpha \\
        e^{i\phi} \cos^2\alpha &\cos^2\alpha \\
    \end{array}\right).
\end{equation}
Hence, the path coherence is
\begin{equation}
    \label{Eq:CohQCRE}
    C_{\text{re}}(\rho_S)=h\left(\tfrac{\cos^2\alpha}{2}\right)-h\left(\tfrac{1+\lambda_{\alpha}}{2}\right),
\end{equation}
where $h(u) \coloneqq -u\log_2{u} - (1-u)\log_2{(1-u)}$ is the binary entropy and
\begin{equation}
    \lambda_{\alpha}\coloneqq\sqrt{2\cos^4\alpha - 2\cos^2\alpha + 1}.
\end{equation}
The predictability is
\begin{equation}
    \label{Eq:PredictQCRE}
    P_{\text{vn}}(\rho_S) = 1 - h\left(\tfrac{\cos^2\alpha}{2}\right).
\end{equation}
Finally, the reduced entropy, in this case, amounts to the entanglement between the system and the BS. Its value is
\begin{equation}
    \label{Eq:EntanQCRE}
    S(\rho_S)=h\left(\tfrac{1+\lambda_{\alpha}}{2}\right).
\end{equation}
Therefore, differently from the QDCE, we see a complementary tradeoff between the components of the CCR, as shown in Fig.~\ref{fig:nr-qcre}. However, as we discuss next, this observation about the QDCE is modified when considering general relativistic effects. Indeed, as will be seen, a tradeoff relation exists in both QDCE and QDCE in this case.

Before concluding, we can also obtain the visibility associated with this setup. Observe that the output state reads
\begin{equation}
    \label{eq:out}
    \ket{\phi_\text{f}} = \frac{1}{\sqrt{2}} \cos{\alpha} \left(\ket{x+} + e^{i\phi} \ket{x-}\right) \ket{\text{in}} + \sin{\alpha} \ket{x+} \ket{\text{out}}.
\end{equation}
With this, it can be shown that this state is associated with precisely the same probability distribution in the ports $D_0$ and $D_1$ as the QDCE~\cite{dieguez2022experimental}. Consequently, both QDCE and QCRE have the same visibility.

\section{Results}
\label{sec:results}

\subsection{General relativistic treatment}
\label{sec:grtreat}

Now, we present our main results and discuss the CCR in the QDCE and QCRE in curved spacetime with a spin-$1/2$ particle. While we focus on these configurations, the general results that we present now are valid regardless of the shape of the interferometers. The geometry of the interferometer will play a more important role in Sec.~\ref{sec:new-lim}, when we make our analysis more concrete by considering systems in the vicinity of Earth.

The study of spin-$1/2$ particles in a generic spacetime requires the use of local reference frames (LRFs)~\cite{terashima2004einstein, Lanzagorta}, which are briefly reviewed in Appendix~\ref{app:boosts}. In this scenario, it turns out that translations in spacetime have an interesting property: besides affecting the description of the particle's momentum, as in a standard relativistic boost, they also introduce a rotation to the system's spin~\cite{silberstein1914theory, thomas1926motion, wigner1939unitary}, known as Wigner rotation. An introduction to the subject is presented in Appendix~\ref{app:wigner}.

This rotation is a function of the spacetime and the momentum of the wave function. In the case of our analysis, we will consider wave packets traveling the paths $|0\rangle$ and $|1\rangle$ in a coherent superposition. As discussed in Appendix~\ref{app:localized-wp}, each of these packets is assumed to be effectively associated with (possibly distinct) momenta $\bar{\mathbf{p}}_0$ and $\bar{\mathbf{p}}_1$. Each path $\eta$ is generally associated with distinct Wigner rotations $D(W(x_\eta, \tau_\eta))$, where $x_\eta$ is the location of the center of mass of the wave packet on path $\eta$ and $\tau_\eta$ is its proper time. 

To start our analysis, consider the QDCE setup depicted in Fig.~\ref{fig:interferometers}(a) in a relativistic scenario, including the presence of curved spacetime. As usual, we assume the system starts in the state\begin{equation}
    \ket{\Psi_\text{i}} = \ket{0} \otimes \ket{\tau} \otimes \ket{\text{BS}},
    \label{eq:relativistic-initial-state}
\end{equation}
where $|\tau\rangle$ refers to the initial state of the particle's spin and $|\text{BS}\rangle \equiv (|\text{in}\rangle + |\text{out}\rangle)/\sqrt{2}$ refers to the second BS, which is in a superposition of being present ($|\text{in}\rangle$) or absent ($|\text{out}\rangle$) of the interferometer path.

Following this, the phase shifter is applied to the system. Then, right after this and before the second BS, the state of the joint system is
\begin{equation}
    \ket{\Psi_\text{QDCE}} = \frac{1}{\sqrt{2}} \left(\ket{0} \ket{\tau_0} + e^{i\phi} \ket{1} \ket{\tau_1}\right) \otimes \ket{\text{BS}},
    \label{eq:qdce}
\end{equation}
where $\ket{\tau_{\eta}} \coloneqq D(W(x_\eta, \tau_\eta)) \ket{\tau}$, with $\eta\in\{0,1\}$. Observe that the state in Eq.~\eqref{eq:qdce} is not constant in this region of the interferometer. Indeed, the states $|\tau_0\rangle$ and $|\tau_1\rangle$ are generally rotated along the path.

From Eq.~\eqref{eq:qdce}, one can see that we have a separable state between the degrees of freedom of the particle and the BS in superposition. However, the spacetime geometry entangles the paths with the spin states. As a result, the reduced density matrix of the paths in superposition, $\rho_S = \Tr_{\text{spin}, \text{BS}} \ket{\Psi_\text{QDCE}}\bra{\Psi_\text{QDCE}}$, is given by
\begin{equation}
     \rho_S = \frac{1}{2} \begin{pmatrix}
           1 & e^{-i \phi} \braket{\tau_1}{\tau_0}  \\
           e^{i \phi} \braket{\tau_0}{\tau_1} & 1
    \end{pmatrix}.
\end{equation}
From this, we see that the CCR should be affected by time dilation since the coherence of $\rho_S$ is affected by the entanglement between the paths and the spin.

\begin{figure}
    \centering
    \includegraphics[width=1\columnwidth]{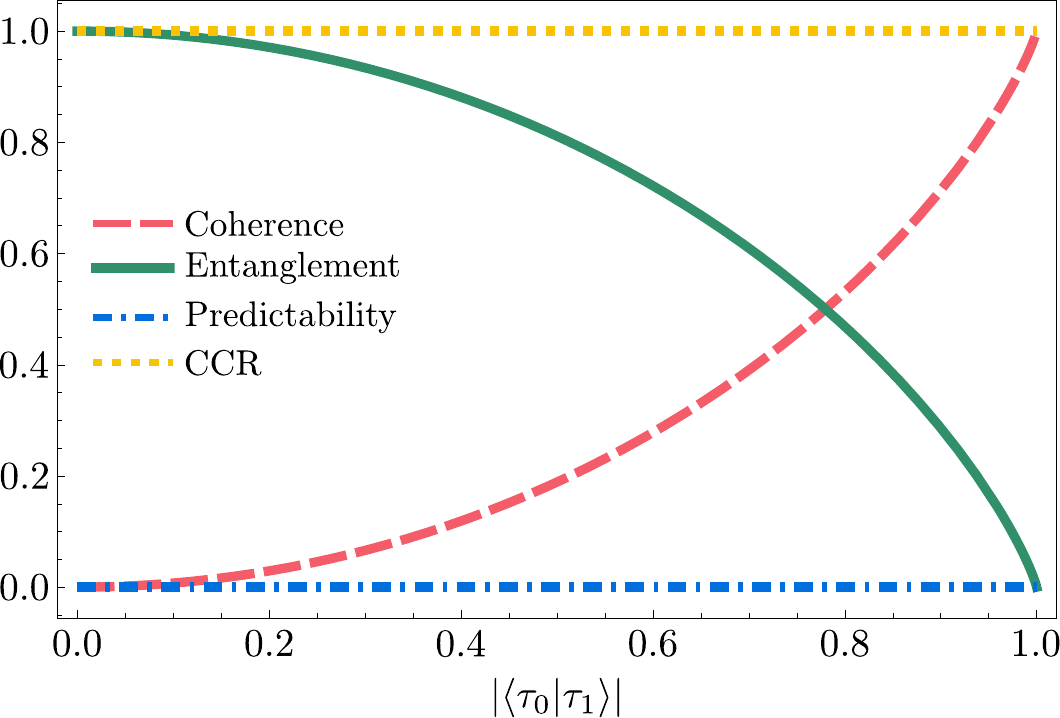}
    \caption{CCR for a particle undergoing the QDCE as a function of the relative Wigner rotations $\abs{\braket{\tau_0}{\tau_1}}$. The graph shows how relativistic effects generally create a tradeoff between coherence and entanglement. This differs from the standard (nonrelativistic) scenario, which is equivalent to the case $\abs{\braket{\tau_0}{\tau_1}} = 1$, where entanglement vanishes. The predictability is not affected by the rotations.}
    \label{fig:rel-qdce}
\end{figure}

To be more precise, we can then compute the elements of the CCR of a system in this region of the interferometer. By direct calculation, it follows that $S(\rho_S)=h(w)$, where $w=\frac{1}{2}(1+\abs{\braket{\tau_0}{\tau_1}})$, and $S(\Phi_Z(\rho_S))=1$. These, in turn, imply that the coherence associated with the paths is
\begin{equation}
    C_{\text{re}}(\rho_S)=1-h\left(\frac{1}{2}(1+\abs{\braket{\tau_0}{\tau_1}})\right),
    \label{eq:c-qdce}
\end{equation}
while the predictability is
\begin{equation}
    P_{\text{vn}}(\rho_S)=0,
    \label{eq:p-qdce}
\end{equation}
and the reduced entropy, which in this case can be associated with the amount of entanglement between the paths and the spin, is
\begin{equation}
    S(\rho_S) = h\left(\frac{1}{2}(1+\abs{\braket{\tau_0}{\tau_1}})\right).
    \label{eq:s-qdce}
\end{equation}
In Fig.~\ref{fig:rel-qdce}, we can see how these quantities change as a function of the relative Wigner rotation between the arms in this setup. Since predictability vanishes, coherence and entanglement are free to oscillate in their full range, in a complementary manner. Depending on the spacetime curvature and the dimensions of the interferometer, multiple periods of these oscillations may take place while the particle traverses the interferometer. As will be seen, this does not happen to be the case in the Newtonian limit for reasonable interferometric sizes.

Finally, we can analyze the visibility of the relativistic QDCE. Right after the second BS (which is in a superposition), the state of the joint system is
\begin{equation}
    \begin{aligned}
        \ket{\Psi_\text{f}} = \; &\frac{1}{\sqrt{2}} \cos\alpha \left(\ket{x+} \ket{\tau_0} + e^{i \phi} \ket{x-} \ket{\tau_1} \right)\ket{\text{in}} \\
        &+ \frac{1}{\sqrt{2}} \sin\alpha \left(\ket{0} \ket{\tau_0} + e^{i\phi} \ket{1} \ket{\tau_1}\right) \ket{\text{out}}.
    \end{aligned}
\end{equation}
Therefore, the probability of detection in $D_{\eta}$, $\eta\in\{0,1\}$, is given by
\begin{equation}
    \mathfrak{p}_{\eta} = \frac{1}{2}\left[1 + (-1)^{\eta} \abs{\braket{\tau_0}{\tau_1}} \cos\phi \cos^2\alpha \right], 
\label{eq:probcur}
\end{equation}
which implies that the visibility is
\begin{equation}
    \mathcal{V} = \abs{\braket{\tau_0}{\tau_1}} \cos^2\alpha = \abs{\cos(\frac{\Theta(1) - \Theta(0)}{2})} \cos^2\alpha,
    \label{eq:viscur}
\end{equation}
where $\Theta(\eta)$ denotes the net Wigner rotation along path $\eta$, which can be completely characterized by the spacetime metric, the path $\eta$, and the initial state of the particle's spin. Observe that now we have a combination of the cosine due to the quantum-controlled beam splitter and the cosine due to the relative Wigner rotation along the paths.

It is noteworthy that Eq.~\eqref{eq:viscur} extends the conclusions that general relativity induces a universal decoherence effect on quantum superposition~\cite{zych2011quantum, zych2012general, pikovski2015universal, zych2016general} to spin particles in two major ways. The first is that we do not restrict ourselves to any specific spacetime. The second is related to the fact that local Wigner rotations, which encode the effect of the spacetime, are purely kinematical and preclude any dynamics of the spin (in particular, we do not impose any special dynamics for the spin).

Now, let us consider the QCRE in a relativistic scenario. The initial state $|\Phi_\text{i}\rangle$ of the system is again given by Eq.~\eqref{eq:relativistic-initial-state} but this time $|\text{BS}\rangle$ refers to the first BS.

\begin{figure*}
    \includegraphics[width=1\textwidth]{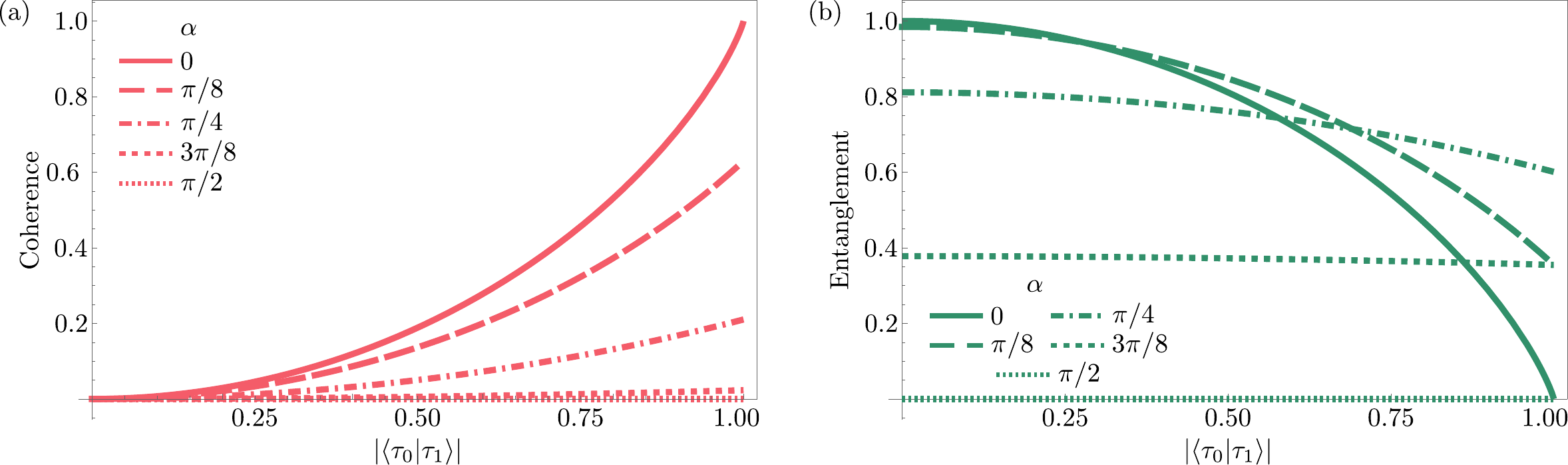}
    \caption{Coherence and entanglement associated with the particle inside the QCRE as a function of the relative state of the particle's spin. When entering the interferometer, $|\langle\tau_0|\tau_1\rangle|=1$. However, this value is generally modified while the particle traverses paths of the interferometer in superposition, introducing a dynamical tradeoff between coherence and entanglement. While predictability is not affected by Wigner rotations, it depends on the parameter $\alpha$, imposing constraints on the range that the values of coherence and entanglement may have. (a) It can be observed that, with an increase in $\alpha$, the amplitude of values of coherence decreases. Indeed, this quantity becomes close to a null constant. (b) The same decrease in the amplitude of values is seen for entanglement. However, it becomes close to a constant that depends on $\alpha$, and is associated with the complement of predictability.}
    \label{fig:rel-qcre}
\end{figure*}

Then, after the interaction with the BS in superposition and passing through the phase shifter, the joint system evolves to
\begin{equation}
    \begin{aligned}
        \ket{\Phi_\text{QCRE}} = &\frac{1}{\sqrt{2}} \cos\alpha \left(\ket{0}\ket{\tau_0} + e^{i \phi} \ket{1}\ket{\tau_1}\right) \ket{\text{in}}  \\
                   &+ \sin\alpha \ket{0} \ket{\tau_0} \ket{\text{out}}.
    \end{aligned}
    \label{eq:psi2w}
\end{equation}
Here, again, the wave packet traveling on each path is effectively associated with a momentum $\bar{\mathbf{p}}_\eta$, where $\eta$ indicates the path. The momenta of distinct paths may have distinct values. Also, like in the QDCE, observe that the states $|\tau_0\rangle$ and $|\tau_1\rangle$ are generally rotated along the path. Thus the above state is not constant in this region of the interferometer. In any case, this expression shows that coherence between the paths is generally affected by time dilation. Indeed, the reduced density matrix of the main system $\rho_S = \Tr_{\text{spin}, \text{BS}} \ket{\Phi_\text{QCRE}}\bra{\Phi_\text{QCRE}}$ is
\begin{equation}
    \rho_S =\frac{1}{2} \begin{pmatrix}
        2- \cos^2\alpha & e^{-i\phi} \cos^2\alpha \braket{\tau_1}{\tau_0}  \\
        e^{i\phi} \cos^2\alpha \braket{\tau_0}{\tau_1} & \cos^2\alpha
    \end{pmatrix}.
\end{equation}

Now, we calculate the components of the CCR of a system in this region of the relativistic QCRE. Observe that the path coherence can be written as
\begin{equation}
    C_{\text{re}}(\rho_S) = h\left(\tfrac{2-\cos^2\alpha}{2}\right)-h\left(\tfrac{1+\lambda_{\alpha'}}{2}\right),
    \label{eq:c-qcre}
\end{equation}
where
\begin{equation}
    \lambda_{\alpha'}\coloneqq\sqrt{\abs{\braket{\tau_0}{\tau_1}}^2\cos^2\alpha + (1 - \cos^2\alpha)^2}.
\end{equation}
Moreover, the predictability is
\begin{equation}
    P_{\text{vn}}(\rho_S) = 1 - h\left(\tfrac{2-\cos^2\alpha}{2}\right).
    \label{eq:p-qcre}
\end{equation}
Finally, the reduced entropy, in this case, can be understood as the amount of entanglement between the paths and the bipartition given by the spin$+$BS. Its value is
\begin{equation}
    S(\rho_S) = h\left(\tfrac{1+\lambda_{\alpha'}}{2}\right).
    \label{eq:s-qcre}
\end{equation}

In Fig.~\ref{fig:rel-qcre}, we can see the tradeoff between these quantities in this setup. Similarly to the QDCE, predictability is insensitive to Wigner rotations. However, it does not vanish in general. Indeed, its value grows with $\alpha$. This imposes a constraint on the interplay between coherence and entanglement, which can no longer oscillate in their full range. For coherence, this means that it decreases its amplitude, becoming closer to a vanishing constant while $\alpha$ grows. The entanglement, however, decreases its oscillation range, approaching constant values that depend on $\alpha$. This constant value also vanishes with $\alpha$ as the predictability value rises to 1. As is the case in the QDCE, the spacetime curvature and the dimensions of the interferometer may allow for little change or for multiple periods of these oscillations before the particle leaves the interferometer. Independently of this, we see that the maximum effect (in magnitude) of Wigner rotations in the CCR takes place when $\alpha=0$.

It is noteworthy that the independence of the predictability $P_{\text{vn}}$ on the spacetime structure is expected. Indeed, from its definition, it can be verified that $P_{\text{vn}}$ would only change if the probability for the system to be found on a given arm of the interferometer changed as well. This cannot be the case since it would, e.g., violate the local probability current, a pillar of quantum theory.

To conclude, we can discuss the visibility associated with this setup. The state of the joint system after the second (standard) BS is
\begin{equation}
    \begin{aligned}
        \ket{\Phi_\text{f}}  = &\frac{1}{\sqrt{2}} \cos\alpha \left( \ket{x+} \ket{\tau_0} + e^{i\phi} \ket{x-} \ket{\tau_1} \right)\ket{\text{in}} \\
              &+ \sin\alpha \ket{x+} \ket{\tau_0} \ket{\text{out}}.
    \end{aligned}
\end{equation}
Then, just like in the QDCE, the probability $\mathfrak{p}_{\eta}$ is given by Eq.~\eqref{eq:probcur} for $\eta\in\{0,1\}$ and, moreover, the visibility of the QCRE amounts to the quantity in Eq.~\eqref{eq:viscur}.

\subsection{Newtonian limit}
\label{sec:new-lim}

To make the results just discussed more concrete, we now consider their Newtonian limit. This limit corresponds to the case of a weak and static gravitational field, such as that in the vicinity of Earth, where the gravitational field can be considered uniform. The spacetime metric in this scenario can be expressed as
\begin{equation}
    ds^2 = - (1 + 2gx)dt^2 + dx^2 + dy^2 + dz^2,
    \label{eq:newtlim}
\end{equation}
where $g = GM/R^2$ is Earth's gravitational acceleration in the origin of the laboratory frame ($x=0$), which is at a distance $R$ from the Earth's center. Moreover, the coordinate $x$ measures the different heights (vertical axis) with respect to the origin of the laboratory frame. Moreover, we associate $z$ with the horizontal axis coordinate. Observe that $x$ is used here with a different meaning than it had in previous parts of this article.

We can then define a tetrad field that represents a static LRF by taking, at each point, the axes $0$, $1$, $2$, and $3$ to be parallel to the directions $t$, $x$, $y$, and $z$, respectively. Then, $e^0_t = (1 + 2gx)^{1/2}$ and $e^1_x  = e^2_y = e^3_z = 1$, while every other component vanishes. The velocity components $u^{\mu}$ of the particle are taken such that the spatial components are constants. In addition, we assume that the speed $u$ of both wave packets is the same and it is constant throughout the entire experiment. Hence, in the LRF, we have $u^0 = -u_0 = \sqrt{1 + u^2}$, $u^1 = u_1 = u_x$, and $u^3 = u_3 = u_z$, where $u^2 = u^2_x +  u^2_z$. Observe that $u^y = 0$ since the particle's dynamics is restricted to the $xz$ plane. Moreover, each wave packet is assumed to never move in a combination of the $x$ and $z$ directions simultaneously since we consider the interferometers with height $h$ and horizontal length $L$ in Fig.~\ref{fig:interferometers}. Also, as discussed in Appendix~\ref{app:spins-clocks}, to maximize the relativistic effects, we assume that $|\tau\rangle$ is any vector that belongs to the intersection between the Bloch sphere and the $xz$ plane.

With this set, it follows that
\begin{equation}
    \abs{\braket{\tau_{0}}{\tau_{1}}} = \abs{\cos(\frac{ (\Theta(1) - \Theta(0)}{2})},
    \label{eq:inner-prod-Thetas}
\end{equation}
where $\Theta(\eta)$ denotes the total Wigner rotation along the path $\eta \in \{0,1\}$ up to the instant location of each wave packet in the laboratory frame, which is given by
\begin{equation}
    \Theta(\eta) = -\gamma \frac{\ell_\eta(t)/u}{\sqrt{1 + 2g(x_0 + \eta h)}},
    \label{eq:Theta-result}
\end{equation}
where $\gamma = g u \sqrt{1 + u^2}$, $x_0$ is the location of the lower horizontal path in the $x$ axis, and $\ell_\eta(t)$ represents the horizontal length traveled by the wave packet until the instant $t$ in the laboratory's clock. Details can be seen in Appendix~\ref{app:new-lim}.

\begin{figure}
    \centering
    \includegraphics[width=\columnwidth]{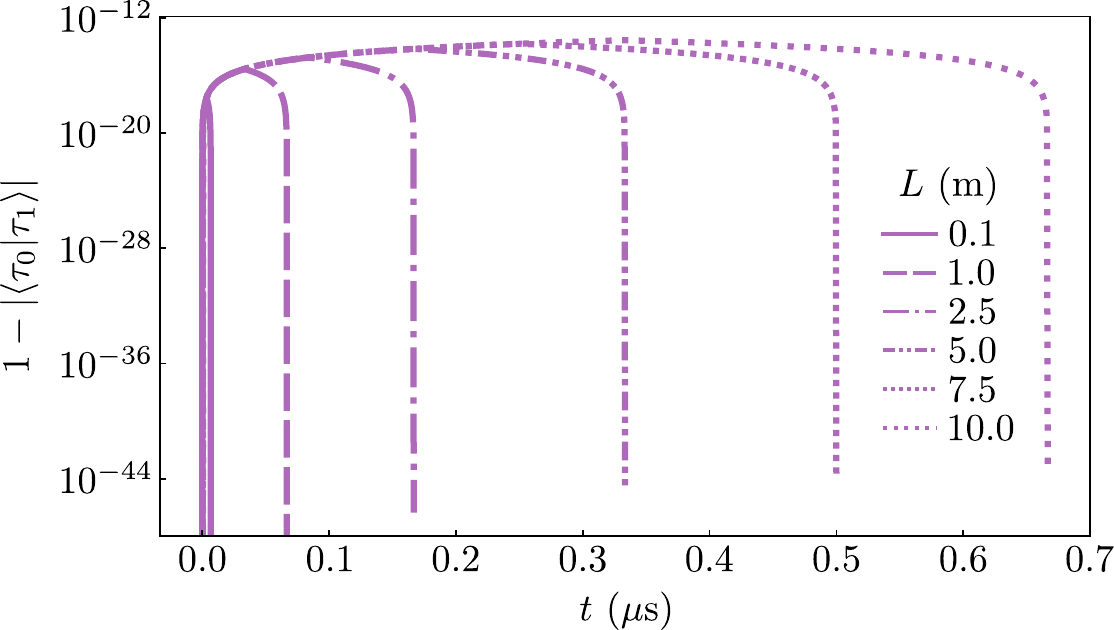}
    \caption{Distinguishability of relative Wigner rotations in the Newtonian limit while a particle encircles squared interferometers. While this quantity, computed from Eqs.~\eqref{eq:inner-prod-Thetas} and \eqref{eq:delta-theta-in}, is small at every location within the interferometer, its value at the instant the particle passes through the second BS is much lower than its maximum value inside. This implies a bigger influence of Wigner rotations in the CCR quantities than may be suggested by the interferometric visibility. It was assumed that $u=10^{-1}c$, $c=3 \times 10^8$ m/s, $g=10$ m/s$^2$, and $x_0=0$.}
    \label{fig:inner-prod-nl}
\end{figure}

We are then ready to compute the elements of the CCR for both the QDCE and QCRE. From Eq.~\eqref{eq:Theta-result}, we obtain
\begin{equation}
    \Theta(1) - \Theta(0) = \gamma \left(\frac{\ell_0(t)/u}{\sqrt{1 + 2gx_0}} - \frac{\ell_1(t)/u}{\sqrt{1 + 2g(x_0+h)}}\right)
    \label{eq:delta-theta-in}
\end{equation}
This expression can be used in Eq.~\eqref{eq:inner-prod-Thetas} to compute the distinguishability of the effects of Wigner rotations on each path. In Fig.~\ref{fig:inner-prod-nl}, the complement of $|\langle\tau_0|\tau_1\rangle|$ is shown as a function of time while a particle traverses squared interferometers with different sizes. We see that, for ``reasonable lengths,'' the distinguishability associated with the action of Wigner rotations on each path is extremely small. This means that these effects cannot be easily measured. In particular, as will be further discussed soon, the visibility, which is related to the values at the moment the particle leaves the interferometer, does not decrease much due to these gravitational effects. However, we observe that, relative to the final values, the spin states associated with the distinct paths are more differentiable in locations inside the interferometer.

We can also show that, in the Newtonian limit, the lengths considered in Fig.~\ref{fig:inner-prod-nl} can be largely increased while keeping the effects of Wigner rotations small. For this end, observe that the maximum change generated by Wigner rotations takes place at an instant $t'$ such that $\ell_0(t')=L$ and $\ell_1(t')=0$. In this case, Eq.~\eqref{eq:inner-prod-Thetas} becomes $\abs{\braket{\tau_{0}}{\tau_{1}}} = \abs{\cos(\gamma L/u)}$, where we have assumed that $x_0=0$. Using the second order approximation $\cos(x) \approx 1- x^2/2$, it holds that
\begin{equation}
    1 - \abs{\braket{\tau_{0}}{\tau_{1}}} = g^2(1+u^2) L^2.
    \label{eq:max-ip-diff}
\end{equation}
In the international system of units, $u \rightarrow u/c$ and $gL \rightarrow gL/c^2$. Then, even if $L$ were of the order of Earth's diameter, i.e., $10^7$ m, the quantity in Eq.~\eqref{eq:max-ip-diff} would be of order $10^{-14}$.

With Eq.~\eqref{eq:inner-prod-Thetas}, we have the needed ingredient to calculate the relevant quantities in the CCR for the QDCE in Eqs.~\eqref{eq:c-qdce}, \eqref{eq:p-qdce}, and \eqref{eq:s-qdce}. Similarly, Eqs.~\eqref{eq:c-qcre}, \eqref{eq:p-qcre}, and \eqref{eq:s-qcre} can be computed, giving the CCR for the QCRE.

Before concluding, we can also discuss the visibility in the Newtonian limit. Since we are interested in the state of the system that leaves the interferometer, we have $\ell_\eta(t) = L$ for both paths $\eta$. Then,
\begin{equation}
    \Theta(1) - \Theta(0) = \gamma \left(\frac{1}{\sqrt{1 + 2gx_0}} - \frac{1}{\sqrt{1 + 2g(x_0+h)}}\right) \Delta T,
\end{equation}
where $\Delta T = L/u$ represents the time interval measured in the laboratory for each packet to travel in the horizontal direction. Using the approximation $(1 + x)^{-1/2} \approx 1 - x/2$, we write
\begin{equation}
    \abs{\braket{\tau_{0}}{\tau_{1}}} = \abs{\cos (\frac{\gamma \Delta V \Delta T}{2})},
    \label{eq:visi}
\end{equation}
where $\Delta V = gh$ is the difference in the gravitational potential between the paths. This result can be replaced in Eq.~\eqref{eq:viscur} in order to compute the visibility of the QDCE and QCRE.

It is noteworthy that Eq.~\eqref{eq:visi} is the same expression obtained for the visibility of the Mach-Zehnder interferometer in a relativistic treatment with the use of local Wigner rotations in the Newtonian limit~\cite{basso2021interferometric}. Then, the visibility of the QDCE and QCRE can be seen as the associated visibility of the Mach-Zehnder interferometer modulated by a function of $\alpha$.

\begin{figure}
    \centering
    \includegraphics[width=\columnwidth]{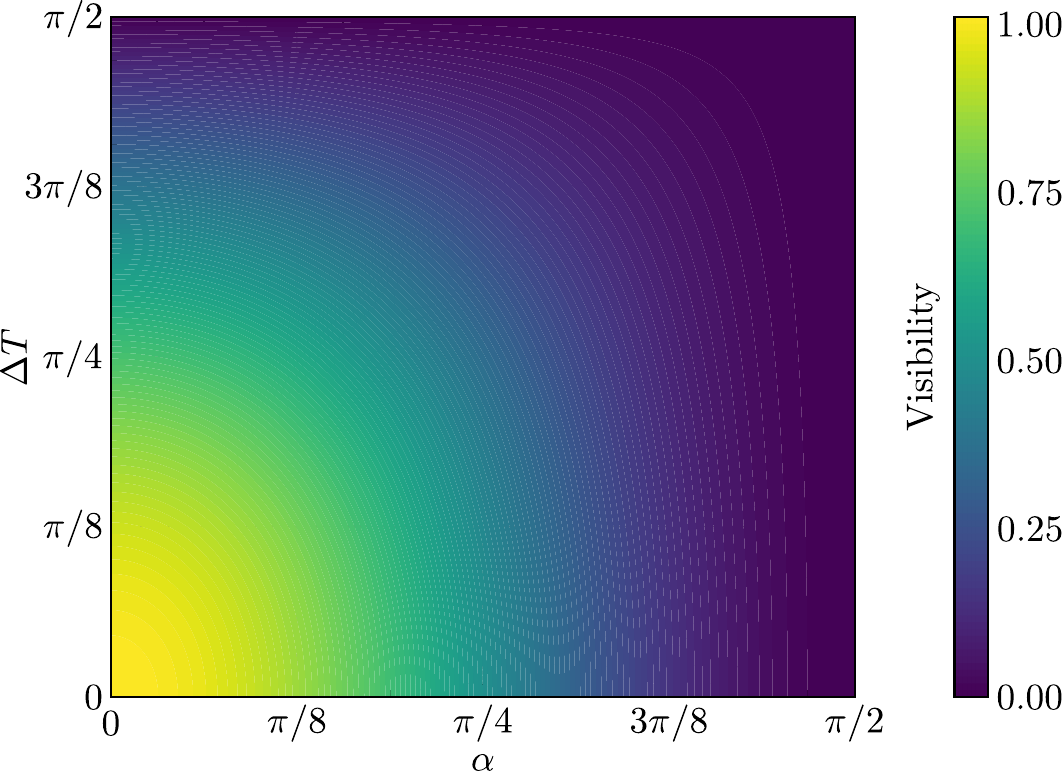}
    \caption{Visibility $\mathcal{V}$ as a function of $\alpha$ and $\Delta T$. The visibility in Eq.~\eqref{eq:viscur} is calculated in the Newtonian limit with $\abs{\braket{\tau_{0}}{\tau_{1}}}$ given by Eq.~\eqref{eq:visi}. For simplicity, it was assumed that $\gamma \Delta V/2 = 1$ s$^{-1}$. Both the controller and the particle's spin store which-path information, contributing to the suppression of the interferometric visibility.}
    \label{fig:nl-visibility}
\end{figure}

In Fig.~\ref{fig:nl-visibility}, we display the dependency of the visibility in the Newtonian limit (with the approximation used in Eq.~\eqref{eq:visi}) as a function of $\alpha$ and $\Delta T$. Observe that both the BS controller and the particle's spin can store at least partial which-way information. As a result, they both contribute to a decrease in the final interferometric visibility. Observe that, in the graphic, $\Delta T$ covers the entire domain needed for the quantity in Eq.~\eqref{eq:visi} to change from 1 to 0. However, in current devices with appropriate scaling such that $\gamma \Delta V/2 = 1$ s$^{-1}$, $\Delta T$ takes values close to 0. This means that the reduction in visibility is minimal.

To give a more concrete notion of the order of magnitude of this effect, we can compare our expressions with the one obtained by Zych \textit{et al.}~\cite{zych2011quantum}, which derives an expression with a form similar to Eq.~\eqref{eq:visi}, albeit different in physical content. Indeed, the authors relied on the internal degree of freedom of the system to not be a spin and to act as a clock. As a result, the analog of the phase in Eq.~\eqref{eq:visi} for this scenario is a vanishing phase.

The reported expression in Ref.~\cite{zych2011quantum} for $|\langle\tau_0|\tau_1\rangle|$ for a spinless particle with an internal clock was $|\cos(\omega \Delta V \Delta T/2)|$, where $\omega$ is the frequency of the clock. We see that the difference between this expression and our expression for a particle with a spin with no dynamics consists of the exchange $\omega \rightarrow \gamma$. If we use the international system of units, this becomes $\omega \rightarrow \gamma/c^2$. Then, even if we consider speeds of the order $u = 10^{-1} c$, we conclude that $\gamma/c^2$ is of order $1/c$. Meanwhile, as discussed in Ref.~\cite{zych2011quantum}, the frequency of accurate clocks can be of order as high as $10^{15}$ Hz. As a consequence, the effect of Wigner rotations in Eq.~\eqref{eq:visi} is much smaller than the effect associated with very precise internal clocks. Therefore, an important conclusion of our analysis is that, in the vicinity of Earth, the results of Ref.~\cite{zych2011quantum} still hold in leading order of magnitude even for spin particles. In this sense, our results extend the analysis of Ref.~\cite{zych2011quantum}.

It is worth emphasizing once more that, despite the extra difficulty of measuring the result in Eq.~\eqref{eq:visi}, it does not rely on the spin of the internal degree of freedom of the system to behave as a clock. This could be an advantage for the experimental verification in some platforms once enough resolution is reached.

Another noteworthy aspect of our results is regarding the discussions in Appendix~\ref{app:spins-clocks}. At least in a first analysis, Wigner rotations present a challenge to approaches that consider clocks as spin degrees of freedom of quantum systems. However, as just seen, in the vicinity of Earth, Wigner rotations do not affect the higher-order dynamics of the free Hamiltonian of the clock, although they may become more relevant in regions of spacetime with more curvature.

\section{Discussion}
\label{sec:discussion}

In this work, we have unveiled tradeoffs between quantities in a CCR for spin particles traversing generalized delayed-choice experiments (QDCE and QCRE). We have presented a nonrelativistic and relativistic analysis of these setups, including a more concrete study in the Newtonian limit. In the nonrelativistic treatment, we have observed that, among the relevant quantities for the CCR, coherence is the only present inside the QDCE. Meanwhile, there is a tradeoff among coherence, entanglement, and predictability inside the QCRE that depends directly on the parameter $\alpha$ that controls the superposition of the BS.

In the relativistic treatment, which involved an arbitrary (potentially curved) spacetime, it was assumed that the particle had an internal degree of freedom, chosen to be a spin for simplicity. Through Wigner rotations, the spin generally becomes correlated with the external degrees of freedom of the particle. This modifies the conclusions drawn in the nonrelativistic scenario, and we have examined these changes in detail.

On the one hand, we have shown that there is now a tradeoff between coherence and entanglement inside the QDCE. The path predictability remains insensitive to Wigner rotations and vanishes. On the other hand, inside the QCRE, there also exists a similar tradeoff between coherence and entanglement. The difference is that, while predictability is not affected by Wigner rotations, it is a function of $\alpha$, as in the nonrelativistic scenario. This adds a constraint in the range of values that coherence and entanglement can take.

Moreover, we have proved that the two interferometers have equal visibility in both nonrelativistic and relativistic treatments. In the relativistic case, the interferometric visibility generally decreases because it is modulated by a quantity associated with the overall effective Wigner rotation during the particle's travel. However, an increase in current measurement precision is needed for the observation of our predictions.

It is worth repeating that, while our results lead to smaller changes to predictions in the vicinity of Earth than previous works in the area, they do not rely on the particle traveling the interferometer having an internal clock. This means that the free Hamiltonian of the internal degree of freedom of the particle considered here may even vanish (as we have assumed for simplicity). In this sense, our study also generalizes (for spin particles) the conclusion from Refs.~\cite{zych2011quantum, zych2012general, pikovski2015universal, zych2016general} that general relativity has a universal decoherence effect on quantum superpositions, once the local Wigner rotation, which encodes the effect of the spacetime, is purely kinematical and precludes any dynamics of the spin. As already mentioned, this is in contrast with the other works, where the coupling between the internal and external degrees of freedom is dynamical, resulting from a global Hamiltonian that couples both degrees of freedom\footnote{Other types of gravitationally induced decoherence have also been considered in the literature. For instance, Ref.~\cite{danielson2023killing} has shown that a superposed massive system will eventually decohere if it is in the vicinity of a Killing horizon (as the event horizon of any stationary black hole).}.

Finally, an interesting research avenue is the study of multipath interferometry in relativistic scenarios. In this context, one could look for special configurations that enhance the effects of Wigner rotations, facilitating their practical observations in the vicinity of Earth. Moreover, establishing how complementarity behaves operationally in a curved spacetime allows further investigations of the performance of tasks based on complementarity and their subsequent adaptation to optimize their use of this quantum resource according to the spacetime structure. This includes, for instance, security protocols, guaranteed by tight entropic-uncertainty relations and entanglement distillation.

\acknowledgments{The authors would like to thank Eliahu Cohen and Augusto C. Lobo for their comments on a previous version of this manuscript, which greatly helped improve the presentation of our results. M.L.W.B. acknowledges financial support from the S\~{a}o Paulo Research Foundation (FAPESP), Grant No.~2022/09496-8. I.L.P. acknowledges financial support from the ERC Advanced Grant FLQuant. P.R.D. acknowledges support from the Foundation for Polish Science (IRAP project, ICTQT, Contract No. MAB/2018/5, co-financed by EU within Smart Growth Operational Programme) and the NCN Poland, ChistEra-2023/05/Y/ST2/00005 under the project Modern Device Independent Cryptography (MoDIC).}

\appendix

\section{Local reference frames and boosts}
\label{app:boosts}

The central object of analysis in this work is particles with an internal structure, which, for simplicity, are taken to be a spin-$1/2$. Since their study in general spacetime requires the use of LRFs, we review them here.

LRFs are defined at each point $x$ of a spacetime $\mathcal{M}$ using a tetrad field (or vielbein), which is composed of 4-vectors $e_a^\mu(x)$, $a = 0,1,2,3$, orthonormal to each other with respect to the pseudo-Riemannian manifold (or spacetime)~\cite{Wald}, i.e., $e^a_{\mu}(x)e_b^{\mu}(x) = \delta^{a}_{b}$ and $e^a_{\mu}(x)e_a^{\nu}(x) = \delta_{\mu}^{\nu}$, where we have used the convention that latin indices refer to coordinates in the LRF, greek indices refer to the general coordinate system defined in the spacetime $\mathcal{M}$, and repeated indices are summed over. Thus, at each point $x \in \mathcal{M}$, the Minkowski metric in the LRF, $\eta_{ab} = \text{diag}(-1,1,1,1)$, and the spacetime metric, $g_{\mu\nu}(x)$, are related to the tetrad field according to~\cite{Nakahara}
\begin{equation}
    \begin{aligned}
        & g_{\mu \nu}(x)e_a^{\mu}(x)e_b^{\nu}(x) = \eta_{ab}, \\
        & \eta_{ab}e^a_{\mu}(x)e^b_{\nu}(x) = g_{\mu \nu}(x).
    \end{aligned}
    \label{eq:metr}
\end{equation}

For the general coordinate system, the lowering and raising of indices are done with the metric $g_{\mu\nu}$ and its inverse $g^{\mu\nu}$, respectively, while the indices in the LRF are lowered by $\eta_{ab}$ and raised by its inverse $\eta^{ab}$. Furthermore, the components of the tetrad field and its inverse transform a tensor in the general coordinate system into one in the local frame and vice versa. Therefore, from Eq.~\eqref{eq:metr}, one can see that the tetrad field incorporates the spacetime curvature information hidden in the metric. In addition, the vielbein $\{e_a^{\mu}(x), a = 0,1,2,3\}$ is a set of four 4-vector fields and transforms under local Lorentz transformations (LLTs) in the local system. Observe that, since $e_0^{\mu}(x)$ is a time-like vector field defined at each point of the spacetime and produces a global time coordinate, it makes the spacetime time orientable~\cite{Wald}. Moreover, the LRF is not unique since it continues to be a LRF under LLTs. Thus, a vielbein representation of a given metric is not defined in a unique manner, and different vielbein connected by LLTs are associated with the same metric tensor~\cite{Lanzagorta}.

Now, we follow the particle's path from one point to another in the curved spacetime $\mathcal{M}$ to construct a representation for LLTs. For this, consider a system with four-momentum $p^{\mu}(x) = m u^{\mu}(x)$ at a certain location $x$, where $m$ is the mass of the particle and $u^{\mu}(x)$ is its four-velocity. Using units such that $c=1$, the momentum satisfies $p^{\mu}(x) p_{\mu}(x) = -m^2$. In the LRF at the point $x$ whose coordinates are $x^a = e^a_{\mu}(x) x^{\mu}$, the momentum of the particle can be written as $p^a(x) = e^a_{\mu}(x) p^{\mu}(x)$. After an infinitesimal interval of proper time $d\tau$ has passed, the particle is found at a new location ${x'}^{\mu} = x^{\mu} + u^{\mu} d\tau$ with momentum $p^a(x') = p^a(x) + \delta p^a(x)$. The variation in momentum can be decomposed into two parts:
\begin{equation}
    \delta p^a(x) = e^a_{\mu}(x) \delta p^{\mu}(x) + \delta e^a_{\mu}(x)p^{\mu}(x),
    \label{eq:momen}
\end{equation}
where the change due to the geometry of spacetime is encoded in $\delta e^a_{\mu}$, while the change $\delta p^{\mu}(x)$ is associated with external non-gravitational forces. The variation $\delta p^{\mu}(x)$ can be written as
\begin{equation}
    \begin{aligned}
        \delta p^{\mu}(x) &= u^{\nu}(x) \nabla_{\nu} p^{\mu}(x) d\tau \\
              &=  -\frac{1}{m}[a^{\mu}(x) p_{\nu}(x) -p^{\mu}(x) a_{\nu}(x)] p^{\nu}(x) d\tau,
    \end{aligned}
\end{equation}
where $a^{\mu} \coloneqq u^{\nu} \nabla_{\nu} u^{\mu}$ is the particle's four-acceleration. Moreover, the variation due to the geometry of spacetime can be written as
\begin{equation}
    \delta e^a_{\mu}(x) = u^{\nu}(x) \nabla_{\nu} e^a_{\mu}(x) d\tau = -u^{\nu}(x) \omega_{\nu b}^{a}(x) e^b_{\mu}(x)d \tau,
\end{equation}
where $\omega_{\nu b}^a \coloneqq e^{a}_{\lambda} \nabla_{\nu} e_{b}^{\lambda} = -e_{b}^{\lambda} \nabla_{\nu} e^{a}_{\lambda} $ is the spin connection~\cite{Chandra}. Substituting these expressions in Eq.~\eqref{eq:momen}, one obtains $\delta p^a(x) = \lambda^a_b(x) p^{b}(x) d\tau$, where
\begin{equation}
    \lambda^a_b(x) \coloneqq u^a(x)a_b(x) - a^{a}(x)u_{b}(x) - u^{\nu}(x) \omega_{\nu b}^a(x).
    \label{eq:def-lambda}
\end{equation}
The tensor $\lambda^a_b(x)$ is interpreted as the infinitesimal LLT since it holds that $p^a(x') = [\delta^a_b + \lambda^a_b(x) d\tau] p^b(x)$~\cite{Lanzagorta}.

\section{Wigner rotations}
\label{app:wigner}

With LRFs and LLTs defined, we follow Refs.~\cite{terashima2004einstein, Lanzagorta} and introduce local Silberstein-Thomas-Wigner rotations \cite{silberstein1914theory, thomas1926motion, wigner1939unitary}, which will be simply referred to as Wigner rotations, as they are more commonly known in the physics community.

The joint state of a particle in an eigenstate of momentum with an internal spin can be denoted by $\ket{p^a(x), \sigma; x^{a}, e^a_{\mu}(x), g_{\mu \nu}(x)}$ to evidence its dependency on the vielbein $e^a_{\mu}(x)$ and $g_{\mu \nu}(x)$ as described from the position $x^a = e^a_{\mu}(x) x^{\mu}$ of the LRF. In this expression, $\ket{\sigma}$ is associated with the particle's internal degree of freedom, i.e., the particle's spin. However, for simplicity, we identify
\begin{equation}
    \ket{p^a(x), \sigma} \equiv \ket{p^a(x), \sigma; x^{a}, e^a_{\mu}(x), g_{\mu \nu}(x)}.
    \label{eq:def-eigen-p}
\end{equation}

Generally, the particle will be in a linear combination of different momentum eigenstates, i.e.,~\cite{Lanzagorta}
\begin{align}
    \ket{\Psi} = \int d \mu(\mathbf{p}) \psi(\mathbf{p})  \ket{\mathbf{p}(x), \sigma}.
    \label{eq:generic-state}
\end{align}
However, as it will become clear, in order to construct the Wigner rotations, it is convenient to take the state of the system to be the one in Eq.~\eqref{eq:def-eigen-p} for a given $p^a(x)$. In this case, it is possible to show that the particle and its internal degree of freedom transform under a spinorial unitary representation of the LLT~\cite{terashima2004einstein, Lanzagorta}, i.e.,
\begin{equation}
    U(\Lambda(x))\ket{p^a(x), \sigma} = \ket{\Lambda p^a(x)} \otimes D(W(x)) \ket{\sigma} \label{eq:unit},
\end{equation}
where $W(x) \coloneqq W(\Lambda(x), p(x))$ is a local Wigner rotation and $D(W(x))$ is a unitary representation of the local Wigner rotation. Therefore, the state $\ket{p^a(x), \sigma}$ at point $x$ is now described as $U(\Lambda(x))\ket{p^a(x), \sigma}$ at point $x'$, and Eq.~\eqref{eq:unit} expresses how the spin rotates locally as the quantum particle moves along its world line.

For an infinitesimal LLT, the corresponding infinitesimal local Wigner rotation can be written as $ W^{a}_b(x) = \delta^a_b + \vartheta^a_b d\tau$~\cite{Kilian}, where
\begin{equation}
    \vartheta^i_j(x) = \lambda^i_j(x) + \frac{\lambda^i_0(x)p_j(x) - \lambda_{j0}(x)p^i(x)}{p^0(x) + m}, \label{eq:wigrot}
\end{equation}
with all other terms vanishing. Therefore, the spin-$1/2$ representation of the infinitesimal local Wigner rotation can be written as
\begin{equation}
    \begin{aligned}
        D(W(x)) &= I_{2 \times 2} + \frac{i}{4} \sum_{i,j,k = 1}^{3} \epsilon_{ijk} \vartheta_{ij}(x) \sigma_k d \tau \\
             &= I_{2 \times 2} + \frac{i}{2} \boldsymbol{\vartheta} \cdot \boldsymbol{\sigma} d\tau,
    \end{aligned}
    \label{eq:wigner}
\end{equation}
where $I_{2 \times 2}$ is the identity matrix and $\sigma_k, \ k =1,2,3$ are the well-known Pauli matrices. Finally, the spin-$1/2$ representation of the local Wigner rotation for a finite interval of proper time is obtained by iterating the expression above, which leads to~\cite{terashima2004einstein}
\begin{equation}
    D(W(x, \tau)) = \mathcal{T}e^{\frac{i}{2}\int_0^{\tau} \boldsymbol{\vartheta} \cdot \boldsymbol{\sigma} d\tau'},
    \label{eq:time}
\end{equation}
where $\mathcal{T}$ is the time-ordering operator.

\section{Spins and clocks}
\label{app:spins-clocks}

As previously explained, the spin rotates locally as the associate quantum particle moves along its world line according to the expression given by Eq.~\eqref{eq:time}. This rotation depends on the momentum of the particle and the spacetime curvature. However, observe that the rotation magnitude is not always proportional to the proper time of a given wave packet. Hence, the spin does act as a clock as a result of these rotations. Nevertheless, it captures the existence of time dilation in some circumstances, as can be concretely seen in Appendix~\ref{app:new-lim}, which explains why we use $\tau$ to denote the state of the particle's spin.

The reason the time-ordering operator $\mathcal{T}$ is needed in Eq.~\eqref{eq:time} is that, generally, the rotations at different points along the particle's trajectory do not commute with each other. This means that, indeed, the spin is generally affected by Wigner rotations regardless of its initial state. However, in some instances, rotations at distinct instants of proper time might commute, i.e., they might be about a single axis. In these cases, particular initial states would not be affected by it. For example, as is the case in Secs.~\ref{sec:grtreat} and~\ref{sec:new-lim}, the Wigner rotation might be a rotation about the $y$ axis, once the motion of the particle inside the interferometer is restricted to a two-dimensional spatial plane. Then, eigenstates of $\sigma_y$ would not be modified and become correlated to the external degree of freedom of the particle. Moreover, while any other initial state could register the rotation in this scenario, a spin's dynamics on the $xz$ plane, i.e., the equator of the $y$ axis, would be optimal, in the sense of allowing for the best distinguishability between states rotated in different manners. Therefore, in this work, we take the initial state of the spin to be any state in the aforementioned equator.

It turns out that this choice is consistent with the prescription to build time states in the Page and Wootters framework~\cite{page1983evolution}, which has also been studied in relativistic and/or gravitational scenarios~\cite{smith2019quantizing, diaz2019history, castro2020quantum, smith2020quantum, giacomini2021spacetime, singh2021quantum, paiva2022non, favalli2022model}. This is a framework for relational dynamics in which time is not an \textit{a priori} parameter. Instead, it is given by a set of states, dubbed time states, from a physical system used as a clock. To construct the set of time states, one first chooses a reference state $|t'\rangle$ in the clock system. The other time states are defined as $|t\rangle \coloneqq  e^{-iH_C (t-t')} |t'\rangle$, where $H_C$ is the Hamiltonian of the clock. However, the set of states $|t\rangle$ should be such that $\int dt \; |t\rangle\langle t| = I_C$, where $I_C$ is the identity operator, which shows that the reference state $|t'\rangle$ should be appropriately chosen. In general, the resulting time operator $T = \int dt \; t |t\rangle\langle t|$ is a positive operator-valued measure, but not necessarily a Hermitian operator~\cite{busch1995operational, smith2019quantizing, hohn2021trinity}. In the case of a clock given by a spin with a Hamiltonian proportional to $\sigma_y$, we see that time states should live in the $xz$ plane, which indeed corresponds to the optimal choice discussed above for the state of the spin in the study in Sec.~\ref{sec:new-lim}.

A question that deserves further investigation is whether Wigner rotations should be considered operational features of time for spin particles revealed by quantum theory or a challenge to the relativistic treatment of clocks as internal degrees of freedom since they introduce modifications to the clock's state in addition to the relativistic time dilation, albeit related to it. In Ref.~\cite{lorek2015ideal}, even though this question is approached from a different perspective, the authors seem to agree with the latter in the case of spinless particles.

\section{Approximation for localized wave packets}
\label{app:localized-wp}

Our brief introduction to Wigner rotations reveals that they depend on the momentum of the particle. Then, consider a particle in a state other than a plane wave, i.e., a generic state as given by Eq.~\eqref{eq:generic-state} with $\psi(\mathbf{p}) \neq \delta(\mathbf{p} - \bar{\mathbf{p}})$ for a constant momentum $\bar{\mathbf{p}}$. In this case, even if the spatial distribution of the particle and its spin start in a separable state, they generally evolve into an entangled state while the particle translates in space.

These exact calculations, however, do not typically lead to simple analytic solutions. Then, it is common to assume a wave packet with a mean centroid $\bar{\mathbf{p}}$ around which the particle momentum is properly distributed, e.g., $\psi_{\sigma}(\mathbf{p}) \propto e^{-(\mathbf{p} - \bar{\mathbf{p}})^2 \xi^2}$ for some positive real constant $\xi$. As a result, in the semiclassical approach, the motion of the particle can be regarded by the motion of its center of mass with momentum $\bar{\mathbf{p}}$. These ideas can be formalized with a Wentzel-Kramers-Brillouin expansion of the exact solution~\cite{Lanzagorta}. Indeed, the analysis just described here corresponds to the zeroth order expansion of the solution in $\hbar$. The centroid $\bar{\mathbf{p}}$ need not be constant throughout the dynamics. Indeed, the result is valid as long as the shape of the wave function is kept approximately unchanged during the experiment~\cite{Wald, Palmer}. These ideas can be extended to the case of packets traveling in a coherent superposition of wave packets such that there is no distortion of the wave packet along the path, where each packet can be associated with different centroid values $\bar{\mathbf{p}}$. A more detailed discussion can be found in Refs.~\cite{Esfahani, Palmer, basso2021effect}.

These remarks play an important role in Sec.~\ref{sec:results}, where we consider a particle traveling an interferometer in a coherent superposition of paths. We assume that the motion of the wave packet traveling on each path can be characterized by its mean centroid momentum, as just discussed. Consequently, each path will be associated with a well-defined time dilation.

\section{Derivations for the Newtonian limit}
\label{app:new-lim}

Given the construction of the tetrads and the conditions described in Sec.~\ref{sec:new-lim}, we now derive the main results presented there. Since the interferometer paths are not geodesic, an external force is necessary to maintain the particle in such a world line. The non-vanishing components of the acceleration due to the external force are
\begin{equation}
    a^0(x) = \frac{g u_x \sqrt{1 + u^2}}{1 + 2gx}, \qquad a^1(x) = \frac{g(1 + u^2)}{1 + 2gx}.
    \label{eq:acce}
\end{equation}

To obtain the infinitesimal LLT, Eq.~\eqref{eq:def-lambda} shows that we first need to compute the spin connection $\omega_{\nu b}^a$. Observe that
\begin{equation}
    \omega_{\nu b}^a = -e_b^\lambda \partial_\nu e^a_\lambda + \Gamma_{\nu\lambda}^\sigma e_\sigma^a e_b^\lambda = \Gamma_{\nu\lambda}^\sigma e_\sigma^a e_b^\lambda,
\end{equation}
where $\Gamma_{\nu\lambda}^\sigma$ is the Christoffel symbol. It can be verified that the only non-zero Christoffel symbols are given by $\Gamma^{x}_{tt}  = g$ and $\Gamma^{t}_{tx} = \Gamma^{t}_{xt} = g / (1 + 2gx)$. Then, by symmetry, we conclude that the non-vanishing components of the spin connection are
\begin{equation}
    \begin{aligned}
        \omega_{t 1}^0 &= \frac{g}{\sqrt{1 + 2gx}}, \\
        \omega_{x 1}^0 &= g\sqrt{1 + 2gx}.
    \end{aligned}
\end{equation}
As a result, the non-zero infinitesimal LLTs are given by
\begin{equation}
    \begin{gathered}
        \lambda^0_1(x) = \frac{g \sqrt{m^2 + p^2}}{m^3(1 + 2gx)} (p^2 - p_x^2) - \frac{\sqrt{1 + 2gx}}{m} p_x, \\
        \lambda^0_3(x) = - \frac{g p_x p_z \sqrt{m^2 + p^2}}{m^3(1 + 2gx)}, \\
        \lambda^1_3(x) = - \frac{g p_z (m^2 + p^2)}{m^3(1 + 2gx)},
    \end{gathered}
\end{equation}
where $p = \sqrt{p^2_x + p^2_z}$. Therefore, the only non-null element of the Wigner rotation in parts of the path is $\vartheta^1_3$. On the horizontal portions of the paths, $p_z=0$ and hence Eq.~\eqref{eq:wigrot} implies that $\vartheta^1_3 = 0$. On the vertical portions of the paths, $p_x=0$ and
\begin{equation}
    \vartheta^1_3(x) = - \frac{g p_z \sqrt{p_z^2 + m^2}}{(1 + 2gx) m^2}.
\end{equation}
From the above expression, we see that the Wigner rotation, when non-null, depends on $g$ and on the height $x$.

The relevant quantity we want to calculate is $|\langle \tau_0| \tau_1\rangle|$, which is the same for both QDCE and QCRE. As a matter of fact, given a spacetime metric, this quantity only depends on the spatial geometry of the interferometer. Using the quantities introduced here, we have
\begin{equation}
    \abs{\braket{\tau_{0}}{\tau_{1}}} = \abs{\expval{e^{-\frac{i}{2}\sigma_y \left(\Theta(1) - \Theta(0)\right)}}{\tau}},
\end{equation}
where
\begin{equation}
    \Theta(\eta) \coloneqq \int_0^{\tau} d\tau' \; \vartheta^1_3(x(\tau';\eta))
\end{equation}
for $\eta\in\{0,1\}$. This leads to Eq.~\eqref{eq:inner-prod-Thetas} if $|\tau\rangle$ is a pure state in the intersection between the Bloch sphere and the $xz$ plane, as it has been assumed. Clearly, only the horizontal regions of each path contribute to $\Theta(\eta)$.

Now, we would like to parametrize the paths in terms of the time $t$ measured by the laboratory frame. Since $-\dd\tau^2 = g_{\mu \nu} \dd x^{\mu} \dd x^{\nu}$, we have
\begin{align}
    \dv{\tau}{t} &= \sqrt{-g_{tt} - g_{ij}\dv{x^i}{t}\dv{x^j}{t}}.
    \label{eq:timedil}
\end{align}
The first term in the square root corresponds to the gravitational time dilation, while the second term corresponds to the special relativistic time dilation. However, if the trajectories are such that the speed of both wave packets is the same and it is constant throughout the entire experiment, as we assume here, there will be no time dilation stemming from special relativity~\cite{zych2011quantum}. As a result, from Eq.~\eqref{eq:timedil}, we are only left with the first term in the square root. Then, since
\begin{equation}
    \Theta(\eta) = \int_0^t dt' \; \frac{d\tau'}{dt'} \; \vartheta^1_3(x(t';\eta)),
\end{equation}
we are led to Eq.~\eqref{eq:Theta-result}.

\bibliography{bibliography.bib}

\end{document}